\documentclass{aa}

\usepackage{graphicx}
\usepackage{txfonts}
\usepackage{natbib}
\bibpunct{(}{)}{;}{a}{}{,} 

\newcommand{\tothe}[1]{\times 10^{#1}}

\defcitealias{Absil06a}{Paper~I}

\begin{document}

\title{Nulling interferometry: performance comparison between Antarctica and other
ground-based sites}

\author{O.~Absil\inst{1,2}\fnmsep\thanks{Marie-Curie EIF Postdoctoral Fellow} \and V.~Coud\'e du
Foresto\inst{3} \and M.~Barillot\inst{4} \and M.~R.~Swain\inst{5}}

\offprints{O. Absil}

\institute{LAOG, Universit\'e Joseph Fourier, CNRS, 414 rue de la Piscine, F-38400 Grenoble, France
\\ \email{olivier.absil@obs.ujf-grenoble.fr}
 \and Institut d'Astrophysique et de G\'eophysique, Universit\'e de Li\`ege, 17 All\'ee du Six
 Ao\^ut, B-4000 Li\`ege, Belgium
 \and LESIA, Observatoire de Paris-Meudon, CNRS, 5 place Jules Janssen, F-92195 Meudon, France
 \and Thales Alenia Space, 100 bd du Midi, F-06156 Cannes La Bocca, France
 \and Jet Propulsion Laboratory, California Institute of Technology, 4800 Oak Grove Drive,
 Pasadena, CA 91109
 }

\date{Received 2 April 2007 / Accepted 28 August 2007}

\abstract
{Detecting the presence of circumstellar dust around nearby solar-type main sequence stars is an
important pre-requisite for the design of future life-finding space missions such as ESA's Darwin
or NASA's Terrestrial Planet Finder (TPF). The high Antarctic plateau may provide appropriate
conditions to perform such a survey from the ground.}
{We investigate the performance of a nulling interferometer optimised for the detection of
exozodiacal discs at Dome~C, on the high Antarctic plateau, and compare it to the expected
performance of similar instruments at temperate sites.}
{Based on the currently available measurements of the turbulence characteristics at Dome~C, we
adapt the GENIEsim software (Absil et al.\ 2006, A\&A 448) to simulate the performance of a nulling
interferometer on the high Antarctic plateau. To feed a realistic instrumental configuration into
the simulator, we propose a conceptual design for ALADDIN, the Antarctic $L$-band Astrophysics
Discovery Demonstrator for Interferometric Nulling. We assume that this instrument can be placed
above the 30-m high boundary layer, where most of the atmospheric turbulence originates.}
{We show that an optimised nulling interferometer operating on a pair of 1-m class telescopes
located 30~m above the ground could achieve a better sensitivity than a similar instrument working
with two 8-m class telescopes at a temperate site such as Cerro Paranal. The detection of
circumstellar discs about 20 times as dense as our local zodiacal cloud seems within reach for
typical Darwin/TPF targets in a integration time of a few hours. Moreover, the exceptional
turbulence conditions significantly relax the requirements on real-time control loops, which has
favourable consequences on the feasibility of the nulling instrument.}
{The perspectives for high dynamic range, high angular resolution infrared astronomy on the high
Antarctic plateau look very promising.}

\keywords{Atmospheric effects -- Instrumentation: high angular resolution -- Techniques:
interferometric -- Circumstellar matter}

\maketitle


\section{Introduction}

Nulling interferometry is considered to be the technique that can enable the spectroscopic
characterisation of the atmosphere of habitable extrasolar planets in the thermal infrared, where
markers of biological activity have been identified \citep{Kaltenegger07}. This is actually the
objective of the Darwin and \mbox{TPF-I} missions studied by ESA and NASA, respectively
\citep{Fridlund04,Beichman99}. While the spectral domain ($6 - 20\,\mu$m, where the atmosphere is
mostly opaque) and required dynamic range ($\sim 10^7$) mandate a space interferometer to achieve
this goal, a ground-based pathfinder might be needed to demonstrate the technique in an operational
context, carry out precursor science, and therefore pave the way for space missions.

One of the main limitations of ground-based nulling interferometers is related to the influence of
atmospheric turbulence. Active compensation of the harmful effects of turbulence requires real-time
control systems to be designed with challenging requirements \citep[][hereafter Paper I]{Absil06a}.
The choice of a good astronomical site with (s)low turbulence is therefore of critical importance.
In this respect, recent studies suggest that the high Antarctic plateau might be the best place on
Earth to perform high-resolution observations in the infrared domain, thanks to its very stable
atmospheric conditions.

The Antarctic plateau has long been recognised as a high-quality site for observational astronomy,
mainly in the context of sub-millimetric and infrared applications, for which the low temperature
and low water vapour content bring a substantial gain in sensitivity. However, the only site that
has been extensively used for astronomy so far is the South Pole, where high wind velocity causes
poor turbulence conditions and thereby prevents high-resolution applications in the near-infrared.
The construction of the French-Italian Concordia station at Dome C (75$^{\circ}$S, 123$^{\circ}$E)
has recently opened the path to new and exciting astronomical studies \citep{Candidi03}. Its main
peculiarity with respect to the South Pole station is that it resides on a local summit of the
plateau (3250~m), where katabatic winds have not yet acquired a significant velocity nor a large
thickness by flowing down the slope of the plateau. For this reason, it is expected that Dome C
could become the best accessible site on the continent, and, given its promising environmental
characteristics, it is worthwhile to investigate its potential for a ground-based nulling
interferometer.


\section{Mission definition}

In order to provide a valid comparison with respect to a temperate site, we chose to study the
potential of the Antarctic plateau in the context of a well specified mission, i.e., the
ground-based Darwin demonstrator that has been identified and studied by ESA at the phase A level
\citep{Gondoin04}. In its original version, GENIE (Ground-based European Nulling Interferometer
Experiment) is conceived as a focal instrument of the VLTI and its science objective is the study
of the exozodiacal dust around nearby solar-type stars like the Darwin targets. Indeed, our
knowledge of the dust distribution in the first few AUs around solar-type stars is currently mostly
limited to the observation of the solar zodiacal disc, a sparse structure of warm silicate grains
10 to 100\,$\mu$m in diameter, which is the most luminous component of the solar system after the
Sun. The presence of similar discs around the Darwin targets (exozodiacal discs) may present a
severe limitation to the Earth-like planet detection capabilities of this mission, as warm
exozodiacal dust becomes the main source of noise if it is more than 20~times as dense as in the
solar zodiacal disc \citep{Beichman06}. On-going interferometric studies are indeed suggesting that
dense exozodiacal discs may be more common than anticipated \citep{Absil06b,DiFolco07}. The
prevalence of dust in the habitable zone around nearby solar-type stars must therefore be assessed
before finalising the design of  the Darwin mission.

Besides its scientific goals, the demonstrator also serves as a technology test bench to validate
the operation of nulling interferometry on the sky.


\section{Atmospheric parameters at Dome C}

Several locations on the Antarctic plateau are expected to provide excellent atmospheric conditions
for high-angular resolution astronomy. Because Dome~C has been extensively characterised during the
past years, it is taken as a reference site for the present study. It might well turn out in the
future that other sites, such as Dome~A or Dome~F, are better suited than Dome~C for the considered
mission. One the one hand, the height of the turbulent ground layer might be thinner than at Dome~C
\citep{Swain06}, while on the other hand, free air seeing\footnote{seeing above the turbulent
ground layer} could be somewhat smaller at Antarctic sites located closer to the centre of the
polar vortex \citep{Marks02}.

    \subsection{Atmospheric turbulence}

Intensive site characterisation at Dome~C has been carried out since the austral summer 2002--03,
with the deployment of several instruments \citep{Aristidi03,Lawrence03}. First, daytime seeing
measurements with a Differential Image Motion Monitor (DIMM) were used to derive a median seeing
value of $1\farcs2$ \citep{Aristidi03}. Later on, using a Multi-Aperture Scintillation Sensor
(MASS) and a Sonic radar (SODAR) in automated mode during wintertime, \citet{Lawrence04} reported a
median seeing of $0\farcs27$. The isoplanatic angle $\theta_0$ and coherence time $\tau_0$ were
also derived from MASS measurements, with average values of $5\farcs7$ and 7.9~msec respectively.
For comparison, the corresponding values at Cerro Paranal are $\theta_0=2\farcs5$ and
$\tau_0=3.3$~msec. These outstanding atmospheric conditions are however valid only above 30~m, as
the SODAR measures the distribution of turbulence in an atmospheric layer comprised between 30 and
900~m above the ground, while the MASS is insensitive to seeing below about 500~m.

These first results suggest that most of the atmospheric turbulence is concentrated in a thin
boundary layer, about 30~m thick. The simultaneous use of two DIMMs at different heights (3~m and
8~m above the ice surface) further confirms this fact, showing that half of the turbulence is
concentrated into the first 5~m above the surface \citep{Aristidi05}. A similar behaviour had
already been reported at the South Pole, where SODAR measurements showed that turbulence was mostly
confined to a boundary layer sitting below 270~m \citep{Travouillon03}. This behaviour can be
explained by the horizontal katabatic wind, whose altitude profile closely matches the turbulence
profile.

In 2005, the first winter-over mission at Dome~C has allowed DIMM measurements and balloon-borne
thermal measurements to be obtained during the long Antarctic night. Preliminary results reported
by \citet{Agabi06} confirm the two-layered structure of atmospheric turbulence at Dome~C. A 36-m
thick surface layer is responsible for 87\% of the turbulence, resulting in a total seeing of
$1\farcs9 \pm 0\farcs5$, while the very stable free atmosphere has a median seeing of $0\farcs36
\pm 0\farcs19$ above 30~m. This value is remarkably similar to the median free air seeing of
$0\farcs32$ reported at South Pole by \citet{Marks99}.

    \subsection{Water vapour seeing}

Another critical parameter for infrared observations is the water vapour content of the atmosphere.
On one hand, it strongly influences the sky transparency as a function of wavelength, and on the
other hand, its temporal fluctuations are an important source of noise for infrared observations.
The water vapour content of the Antarctic atmosphere has been measured at South Pole by radiosonde,
giving an exceptionally low average value of 250~$\mu$m during austral winter
\citep{Chamberlin97,Bussmann05}, where temperate sites typically have a few millimetres of
precipitable water vapour (PWV). This is mainly due to the extreme coldness of the air, with a
ground-level temperature of about $-61^{\circ}$C (212~K) at the South Pole during
winter,\footnote{Dome~C is even colder during winter, with an average ground temperature of
$-65^{\circ}$C (208~K).} which induces a low saturation pressure for water vapour. The winter time
PWV at Dome C is estimated to be between 160~$\mu$m \citep{Lawrence04b} and 350~$\mu$m
\citep{Swain06b}.

\begin{table*}[!t]
\begin{center}
\caption{Water vapour seeing at three astronomical sites. The Fried parameter $r_0$ is given at
500~nm, adapted from the values of \citet{Racine05}. References for the PWV data are given in the
last column. The standard deviation of water vapour seeing at Dome~C is deduced from the data at
Cerro Paranal and Mauna Kea following Equation~\ref{eq:pwv}. Water vapour seeing was measured on a
baseline of 66\,m (resp.\ 100\,m) at Paranal (resp.\ Mauna Kea), and we assume that our estimate at
Dome~C is valid for a similar range of baseline lengths.} \label{tab:pwv}
\begin{tabular}[h]{ccccc} \hline         Site     &  $r_0$  & $\langle {\rm PWV}
\rangle$ & $\sigma_{\rm PWV}$ & References
\\ \hline Cerro Paranal & 14.5~cm & 3~mm       & 27~$\mu$m  & \citet{Meisner02}
\\        Mauna Kea     & 17.8~cm & 1.6~mm     & 11~$\mu$m  & \citet{Colavita04}
\\        Dome C        & 38.2~cm & 0.25~mm    & 1~$\mu$m & \citet{Bussmann05}
\\ \hline
\end{tabular}
\end{center}
\end{table*}

The very low water vapour content of the Dome~C atmosphere has an important advantage in the
context of high-precision infrared interferometry: longitudinal dispersion, created by the
fluctuations of the water vapour column density above the telescopes \citep{Colavita04}, is greatly
reduced with respect to temperate sites. The standard deviation of PWV ($\sigma_{\rm PWV}$) can be
estimated at Dome~C assuming that water vapour seeing follows the same statistics as the piston, as
suggested by \citet{Lay97}. In that case, the standard deviation of PWV fluctuation depends on the
Fried parameter $r_0$ viz.\ $\sigma_{\rm PWV} \propto r_0^{-5/6}$ according to \citet{Roddier81}.
Assuming that $\sigma_{\rm PWV}$ is also proportional to $\langle {\rm PWV} \rangle$, the average
PWV content, its value at Dome~C can then be obtained by means of a comparison with the data
obtained at temperate sites:
\begin{equation}
\sigma_{\rm PWV} ({\rm DC}) = \frac{\langle {\rm PWV} \rangle_{\rm DC}} {\langle {\rm PWV}
\rangle_i} \left( \frac{r_{0,\,{\rm DC}}}{r_{0,\,i}} \right)^{-5/6} \sigma_{\rm PWV} (i) \; ,
\label{eq:pwv}
\end{equation}
where $i$ represents a (well-characterised) temperate site. The application of this relation with
the atmospheric parameters of either Cerro Paranal or Mauna Kea taken as a reference gives very
similar estimates of 1.0 and 0.91~$\mu$m for $\sigma_{\rm PWV}$ at Dome~C (see
Table~\ref{tab:pwv}), using an average PWV content of 250~$\mu$m for Dome~C. For this calculation,
we assumed a seeing of $0\farcs27$, which is valid only above 30~m \citep{Lawrence04}. Using this
value is recommended in the present case for two reasons: on one hand, telescopes are contemplated
to be placed above the turbulent ground layer, and on the other hand, the study of
\citet{Bussmann05} shows that most of the PWV is concentrated between 200~m and 2~km above the
ground, so that water vapour seeing is suspected to be only weakly affected by the ground layer.

    \subsection{Atmospheric transmission and sky brightness}

Another benefit from the low water vapour content is to widen and improve the overall transmission
of the infrared atmospheric windows. \citet{Lawrence04b} shows that the $K$ band is extended up to
2.5~$\mu$m and the $L$ band from 2.9 to 4.2~$\mu$m. The transmission of the $M$ band around
5~$\mu$m is also significantly improved.

The infrared sky brightness is also partially determined by the water vapour content, which affects
its wavelength-dependent emissivity factor. The other parameter influencing the sky emission is its
effective temperature, which depends on the altitude of the main opacity layer at a given
wavelength. The effective temperature above South Pole has been measured by \citet{Chamberlain00}
in the mid-infrared, with values ranging from 210~K to 239~K depending on wavelength. Most of the
winter sky background emission is in fact assumed to emanate from an atmospheric layer just above
the temperature inversion layer, located between 50 and 200~m at Dome~C
\citep{Chamberlain00,Lawrence04b}. The atmospheric temperature at this altitude is about 230~K in
wintertime \citep{Agabi06}. As a result of both low temperature and low emissivity, the sky
background is exceptionally low in Antarctica. The measurements obtained at South Pole show that is
it reduced by a factor ranging between 10 and 100 in the infrared domain with respect to temperate
sites. The largest gain in sensitivity for astronomical observations is expected to arise in the
$K$, $L$ and $M$ bands. It is estimated that 1-m class telescopes at Dome~C would reach almost the
same sensitivity as 8-m class telescopes at a temperate site at these wavelengths.

The atmospheric parameters discussed in this section are summarised in Table~\ref{tab:atmoparam}.

\begin{table}[t]
\begin{center}
\caption{Atmospheric parameters adopted for the performance simulation of ALADDIN at Dome~C,
assuming that the instrument is located about 30~m above the ground level. The equivalent wind
speed is the wind speed integrated across the whole turbulence profile (above 30~m).}
\label{tab:atmoparam}
\begin{tabular}{cc}
\hline \multicolumn{2}{c}{Atmospheric parameters}
\\ \hline Fried parameter $r_0$ at 500~nm & 38~cm
\\        Equivalent seeing & $0\farcs27$
\\        Coherence time $\tau_0$ & 7.9~msec
\\        Equivalent wind speed & 15~m/s
\\        Outer scale $L_{\rm out}$ & 100~m
\\        Sky temperature & 230~K
\\        Ambient temperature at $h=30$~m & 230~K
\\        Mean PWV & 250~$\mu$m
\\        rms PWV & 1~$\mu$m
\\        Pressure & 640~mbar
\\ \hline
\end{tabular}
\end{center}
\end{table}


\section{The ALADDIN nulling interferometer concept}

To provide plausible inputs in terms of instrumental parameters for the performance simulation, a
conceptual design is needed for our Antarctic nulling interferometer. The main design guidelines
that have been adopted are the minimisation of the number of open air reflections, the preservation
of the full symmetry between the two beams and the optimisation of the range of baselines for
typical Darwin target stars. Another critical guideline to benefit from the outstanding free air
seeing is to place the instrument above the boundary layer (i.e., about 30~m above the ground at
Dome C). The following sections describe a practical concept of a nulling interferometer dedicated
to exozodiacal disc detection, which follows these recommendations without pretending to be
optimal. This concept is referred to as ALADDIN, the Antarctic $L$-band Astrophysics Discovery
Demonstrator for Interferometric Nulling.

    \subsection{The interferometric infrastructure}

\begin{figure*}[t]
\centering
\includegraphics[width=18cm]{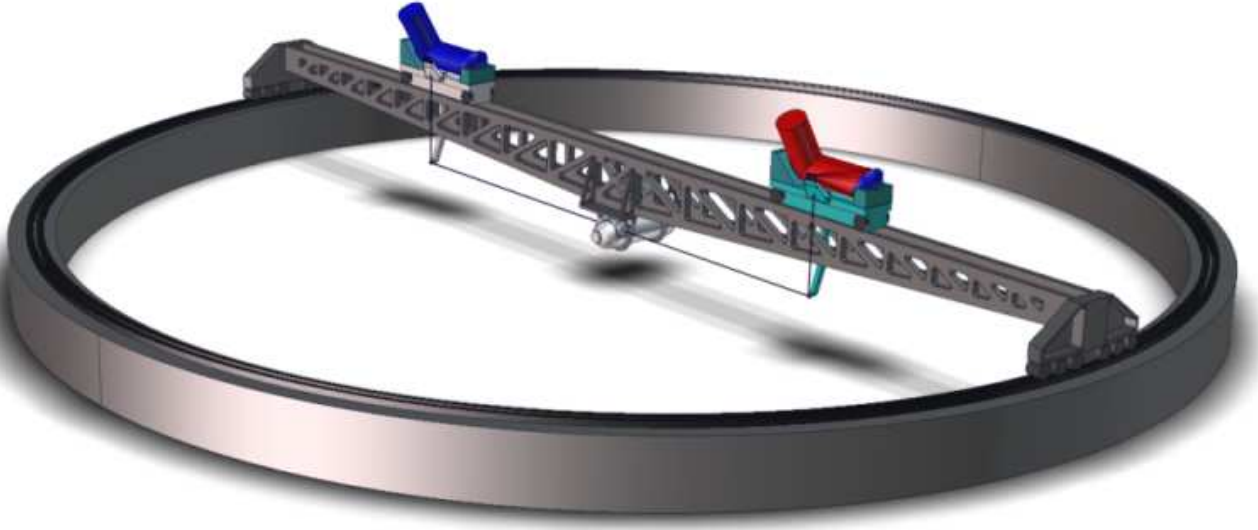}
\caption{Overview of the ALADDIN infrastructure. A 40-m rotating truss bearing the siderostats is
mounted on a 30-m high structure (not represented). The light beams collected by the two
siderostats are fed into off-axis telescopes and routed towards the nulling instrument cryostat by
fixed relay optics (only five reflections outside the cryostat).} \label{fig:aladdin}
\end{figure*}

The concept proposed here consists in a 40~m long rotating truss installed on top of a 30~m tower,
and on which are placed two moveable siderostats feeding off-axis telescopes
(Fig.~\ref{fig:aladdin}). Such a design has two main advantages: first, thanks to the moveable
siderostats, the baseline length can be optimised to the observed target and second, thanks to the
rotating truss, the baseline can always be chosen perpendicular to the line of sight so that
neither long delay lines nor dispersion correctors are needed. Moreover, polarisation issues, which
are especially harmful in nulling interferometry \citep{Serabyn01}, are mitigated by this fully
symmetric design. The available baseline lengths range from 4 to 30~m and provide a maximum angular
resolution of 10~mas in the $L$ band. This is largely sufficient to study the habitable zones
around Darwin/TPF-I candidate targets, since they are typically separated by a few tens of
milliarcseconds from their parent star \citep{Kaltenegger06}.

\begin{table}[t]
\begin{center}
\caption{Summary of the instrumental parameters assumed for the performance simulation of ALADDIN.
The throughput and emissivity are directly computed from the baseline instrumental design, which is
based on a simplified version of the GENIE instrument.} \label{tab:instruparam}
\begin{tabular}[h]{cc}
\hline \multicolumn{2}{c}{Instrumental parameters}
\\ \hline Baselines & $4-30$~m
\\        Telescope diameter & 1~m
\\        Number of warm optics & 5
\\        Warm optics temperature & 230~K
\\        Warm throughput & 80\%
\\        Warm emissivity & 20\%
\\        Number of cold optics & 15
\\        Cryogenic temperature & 77~K
\\        Cold throughput & 10\%
\\        Science waveband & $3.1-4.1$~$\mu$m ($L$)
\\        Fringe sensing waveband & $2.0-2.4$~$\mu$m ($K$)
\\        Tip-tilt sensing waveband & $1.15-1.3$~$\mu$m ($J$)
\\ \hline
\end{tabular}
\end{center}
\end{table}

For the baseline version of the ALADDIN design shown in Fig.~\ref{fig:aladdin}, the diameter of the
siderostats has been set to 1~m, which is expected to provide similar performance to 8-m class
telescopes at a temperate site. Only five reflections are required to lead the light from the sky
down to the instrument, which is accommodated under the rotating truss, at the rotation centre. All
relay mirrors are at ambient temperature, i.e., about 230~K at that altitude during wintertime
\citep{Agabi06}. Note that an alternative design, where the instrument is placed on the ground, was
introduced earlier \citep{Barillot06}. In the latter version the cryostat does not need to be
rotated with the truss and remains fully static, at the cost of a more complex optical train which
enables symmetric de-rotation of the beams and preservation of the polarisation. The harmful
influence of ground-layer seeing is then mitigated by propagating compressed beams about 40~mm in
diameter, i.e., smaller than the typical Fried parameter in the ground layer.

    \subsection{The nulling instrument}

The ALADDIN interferometer feeds a nulling instrument whose design is directly inherited from
GENIE, a nulling instrument originally designed to be installed at ESO's Very Large Telescope
Interferometer (VLTI) on top of Cerro Paranal. Using a common base design has the advantage of
improving the comparative value of the performance simulations.  Indeed, ALADDIN is foreseen to
operate in the same wavelength regime, the $L$ band (ranging from 2.8 to 4.2~$\mu$m at Dome C),
which is very appropriate to investigate the inner region of extrasolar zodiacal discs. The whole
nulling instrument is assumed to be enclosed in a cryostat, in order to improve its overall
stability and to mitigate the influence of temperature variations between seasons at the ground
level (the mean temperature during the austral summer is about 40$^{\circ}$C higher than during
winter, while the instrument should be usable during the whole year). The lower temperature of the
optics inside the cryostat (77~K) also further decreases the background emission produced by the
instrument. Since (as is shown below) two subsystems needed for GENIE are no longer needed for the
Antarctic version, the instrument is expected to be smaller and therefore easier to enclose into a
cryostat.

\begin{figure}[t]
\centering \resizebox{\hsize}{!}{\includegraphics{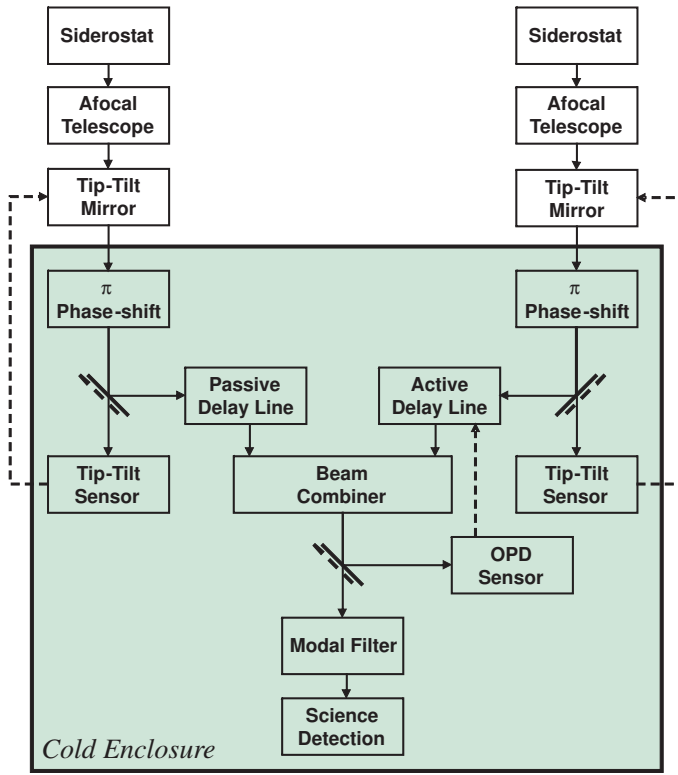}} \caption{Block diagram of the
ALADDIN optical architecture. Feed-back signals driving the control of the tip-tilt and OPD
parameters are shown in dashed lines. The entrance windows of the cold enclosure are not
represented, for simplicity.} \label{fig:block}
\end{figure}

The ALADDIN instrument comprises the same basic functionalities as GENIE (fringe tracking, tip-tilt
correction, phase shifting, beam combination, modal filtering, spectral dispersion and detection),
except for two critical control loops that are not needed any more. As demonstrated in
Section~\ref{sec:aladdinperfo}, ALADDIN can on the one hand be operated without any dispersion
correction thanks to the rotating baseline and to the very low water vapour seeing, provided that
the observing waveband is restricted to the $3.1 - 4.1$~$\mu$m region, while on the other hand,
real-time intensity control is not required any more since the size of the collectors is
significantly smaller than the Fried parameter in the $L$ band ($r_{0, L} \simeq 4$~m). A block
diagram of the optical path and control system of ALADDIN is shown in Fig.~\ref{fig:block}. Most
optical functions are kept at low temperature inside a vacuum enclosure. The optical arrangement
has been significantly simplified with respect to the original VLTI/GENIE design:
\begin{itemize}
\item
The two-mirror afocal telescopes are off-axis. Thanks to the narrow field-of-view, high wavefront
quality is expected.
\item
Tip-tilt correction is performed at the level of the collecting telescopes assemblies, so that the
optical paths downstream are kept identical whatever the baseline and orientations of the
siderostats and structural beam.
\item
The achromatic $\pi$ phase-shift is achieved geometrically, by means of opposite periscopes.
\item
The beam splitters shown in Fig.~\ref{fig:block} are actually dichroic beam splitters, which
separate the signal between the science wave band and the tip-tilt and OPD sensing wave bands.
\item
Optical delay lines are of the short stroke/high accuracy kind, since long stroke is not necessary
in the rotating beam architecture. Their design is expected to be greatly simplified with respect
to usual delay lines: one-stage actuators based on linear piezoelectric motors translating a small
and light plane mirror are expected to be sufficient.
\item
The preferred beam combiner arrangement is the Modified Mach-Zehnder \citep[MMZ,][]{Serabyn01}.
\item
The modal filter is a single-mode optical fibre. Fluoride glass fibres are appropriate for
ALADDIN's science wavelengths.
\item
OPD detection may be achieved downstream the beam combiner, by means of an ABCD algorithm, provided
that one of the two nulled outputs of the MMZ receives a $\pi/2$ phase shift. Alternately, the
separation between OPD sensor and science bands may be implemented upstream the beam combiner and a
second beam combiner accommodated for the OPD measurement. The latter option, which was the
baseline for the GENIE instrument, has been used for performance estimation to provide a fair
comparison with GENIE.
\end{itemize}
The control system involves three control loops only, respectively dedicated to the stabilisation
of one OPD and two tip-tilt parameters. They are expected to be operated at lower repetition
frequencies than at a temperate site thanks to the slowness of atmospheric turbulence, which
represents a significant simplification. The control loops are based on conventional and separated
PID controllers involving separated sensors and actuators. The location of the tip-tilt mirrors in
the output pupil of the telescopes ensures proper uncoupling between tip-tilt actuation and OPD.


\section{Performance study at Dome~C} \label{sec:aladdinperfo}

In order to evaluate the performance of ALADDIN, we use the GENIE simulation software (GENIEsim),
which performs end-to-end simulations of ground-based nulling interferometers with a system-based
architecture. All the building blocks and physical processes included in GENIEsim are described in
detail in \citetalias{Absil06a}. They include the simulation of astronomical sources (star,
circumstellar disc, planets, background emission), atmospheric turbulence (piston, longitudinal
dispersion, wavefront errors, scintillation), as well as a realistic implementation of closed-loop
compensation of atmospheric effects by means of a fringe tracking system and of a wavefront
correction system. The output of the simulator basically consists in time series of photo-electrons
recorded by the detector at the two outputs of the nulling beam combiner (constructive and
destructive outputs). Various information on the sub-systems are also available on output for
diagnostic. Routines dedicated to the post-processing of nulling data are also included, as
described in \citetalias{Absil06a}. GENIEsim is written in the IDL language. It has been originally
designed to simulate the GENIE instrument at the VLTI interferometer, and has been extensively
validated in that context either by comparison with on-sky data when available (e.g., MACAO and
STRAP for adaptive optics, FINITO for fringe tracking) or by comparison with performance
estimations carried out by industrial partners during the GENIE phase A study.

\begin{table*}[t]
\begin{center}
\caption{Control loop performance and optimum repetition frequencies (0~Hz means that no control
loop is used) as simulated on a 100~sec observation sequence, for the GENIE instrument working on
the 8-m Unit Telescopes (UT) at the VLTI (results taken from \citetalias{Absil06a}) and the ALADDIN
instrument at Dome~C. The observations are carried out in the $L$ band for a Sun-like G2V star
located at 20~pc using either the 47-m UT2-UT3 baseline at the VLTI (waveband: 3.5--4.1~$\mu$m) or
a baseline length of 20~m for ALADDIN (waveband: 3.1--4.1~$\mu$m). The goal performance for
exozodiacal disc detection discussed in \citetalias{Absil06a} appears in the last column. The total
null is the mean nulling ratio including both the geometric and instrumental leakage contributions.
The last line gives the standard deviation of the instrumental nulling ratio for this 100~sec
sequence (note that we give here the standard deviation of the mean instrumental nulling ratio
computed on the whole time sequence, which is more representative than the frame-to-frame deviation
presented in \citetalias{Absil06a}).} \label{tab:loopperf}
\begin{tabular}[h]{cccccc}
\hline              &\multicolumn{2}{c}{GENIE -- UT}&   \multicolumn{2}{c}{ALADDIN} &
\\                  &   Worst case  &   Best case   &   Worst case  &   Best case   &   Goal
\\ \hline Piston    & 17~nm @ 20~kHz&6.2~nm @ 13~kHz&  14~nm @ 3~kHz&  10~nm @ 2~kHz& $<4$~nm
\\ Inter-band disp. & 17~nm @ 200~Hz&4.4~nm @ 300~Hz& 7.0~nm @ 0~Hz & 7.0~nm @ 0~Hz & $<4$~nm
\\ Intra-band disp. &4.1~nm @ 200~Hz&1.0~nm @ 300~Hz& 7.4~nm @ 0~Hz & 7.4~nm @ 0~Hz & $<4$~nm
\\         Tip-tilt &11~mas @ 1~kHz &11~mas @ 1~kHz &  9~mas @ 1~kHz&  9~mas @ 1~kHz&(see intensity)
\\Intensity mismatch&   4\% @ 1~kHz &   4\% @ 1~kHz &  1.2\% @ 0~Hz &  1.2\% @ 0~Hz & $<1$\%
\\ \hline Total null&$9.7\tothe{-4}$&$6.2\tothe{-4}$&$2.9\tothe{-4}$&$2.2\tothe{-4}$& $f$(baseline)
\\Instrumental null &$5.0\tothe{-4}$&$1.5\tothe{-4}$&$2.0\tothe{-4}$&$1.3\tothe{-4}$& $10^{-5}$
\\         rms null &$4.5\tothe{-6}$&$2.0\tothe{-6}$&$5.0\tothe{-6}$&$3.5\tothe{-6}$& $10^{-5}$
\\ \hline
\end{tabular}
\end{center}
\end{table*}

Thanks to the versatility of the simulator, only a few input parameters have to be changed to
switch from the original configuration (GENIE at Cerro Paranal) to ALADDIN at Dome~C. These changes
include the atmospheric transmission \citep{Lawrence04b}, as well as the atmospheric and
instrumental parameters listed in Tables~\ref{tab:atmoparam} and~\ref{tab:instruparam}. It must be
noted that the ALADDIN performance can be modelled with greater confidence than in the case of
GENIE as it does not rely on the nominal performance of an external system such as the VLTI.
Furthermore, the performance should remain similar across most of the Antarctic plateau, as free
air seeing is not expected to change drastically for sites located within the polar vortex. The
only requirement is then to adapt the height of the structure on which the instrument is placed. In
this regard, Dome~C might not be the best possible site, as the boundary layer is suspected to be
about 10~m thinner at Dome~F \citep{Swain06}.

As in the case of GENIE, the performance is measured in terms of sensitivity to faint exozodiacal
dust clouds. We assume that these dust clouds follow the same density and temperature distribution
as in the solar system \citep{Kelsall98}, except for a global density scaling factor. To account
for this, we introduce the unit {\em zodi}, which corresponds to the global dust density in our
local zodiacal cloud.

    \subsection{Control loop performance}

Because dispersion and intensity control loops are not expected to be required in the case of
ALADDIN, we have disabled these two loops in the GENIEsim software when simulating the residual
atmospheric turbulence at beam combination. The simulation results are presented in
Table~\ref{tab:loopperf}, where the absence of dispersion and intensity control is represented by a
0~Hz repetition frequency. For these simulations, we have used two different assumptions on the
atmospheric turbulence characteristics. The {\em worst case scenario} does not take into account
the effect of pupil averaging, which is expected to reduce the power spectral density (PSD) of
piston and dispersion at high frequencies \citep{Conan95}. This scenario thereby assumes a
logarithmic slope of $-8/3$ at high frequencies for the PSD of these two quantities. Conversely,
the {\em best case scenario} takes into account the effect of pupil averaging at high frequencies
where it produces a $-17/3$ logarithmic slope. The rationale for introducing the worst-case
scenario is that the $-17/3$ slope has never been observed to our best knowledge (most probably due
to instrumental limitations), while spurious instrumental effects might potentially increase the
high-frequency content of piston. It must be noted that the PSDs of higher order Zernike modes
(tilt and above) remain the same in both scenarios and take into account pupil averaging.

The results listed in Table~\ref{tab:loopperf} confirm that two critical control loops (dispersion
and intensity control) are not required any more: the input atmospheric perturbations for these two
quantities are either well below other contributions (e.g., piston) or marginally compliant with
the goal performance taken from \citetalias{Absil06a}.\footnote{Note that the strength of
dispersion decreases for shorter baselines, and is only about 3\,nm rms for a 4-m baseline.} A
second important conclusion is that, in order to reach a residual piston similar to that of GENIE,
fringe tracking can be carried out at a much lower frequency (about 3~kHz instead of 20~kHz). The
technical feasibility of the instrument directly benefits from these two features. Finally, it must
be noted that the two models for atmospheric turbulence provide similar results. There are two
reasons for this: the actual shape of the power spectral density has no influence on the global
fluctuation of the quantities that are not subject to real-time control, and the cut-off frequency
at which the effect of pupil averaging becomes important is significantly higher than in the case
of GENIE due to the reduced pupil diameter \citepalias[see][]{Absil06a}.

Despite its smaller collecting area, the overall performance of ALADDIN in terms of instrumental
null is slightly improved with respect to GENIE's, by a factor up to~2.5 in the worst-case
scenario. However, the mean instrumental nulling ratio achieved by ALADDIN is still a factor $\sim
10$ above the performance required to detect 20-zodi discs without calibrating the instrumental
response. This shows that, as in the case of GENIE, the calibration of instrumental stellar leakage
will be mandatory to approach the goal sensitivity of 20~zodi.

    \subsection{Estimated sensitivity}

Using the parameters of Tables~\ref{tab:atmoparam} and~\ref{tab:instruparam}, we have simulated the
detection performance of ALADDIN for exozodiacal discs. The simulations take into account the same
calibration procedures as discussed in \citetalias{Absil06a} in the context of GENIE, i.e.,
background subtraction, geometric leakage calibration and instrumental leakage calibration. Four
hypothetic targets, representative of the Darwin star catalogue, have been chosen for this
performance study: a K0V at 5~pc, a G5V at 10~pc, a G0V at 20~pc and a G0V at 30~pc. The
integration time has been fixed to 30~min as in the case of GENIE. Unless specified, we have
assumed a typical uncertainty $\Delta\theta_{\ast}$ of 1\% on the diameters of the target stars
\citepalias[see][]{Absil06a} and we have used the worst case scenario for atmospheric turbulence
with the $-8/3$ logarithmic slope of the power spectra at high frequencies. As demonstrated in
Table~\ref{tab:loopperf}, using the best case scenario would not significantly change the final
results.

\begin{figure}[t]
\centering \resizebox{\hsize}{!}{\includegraphics{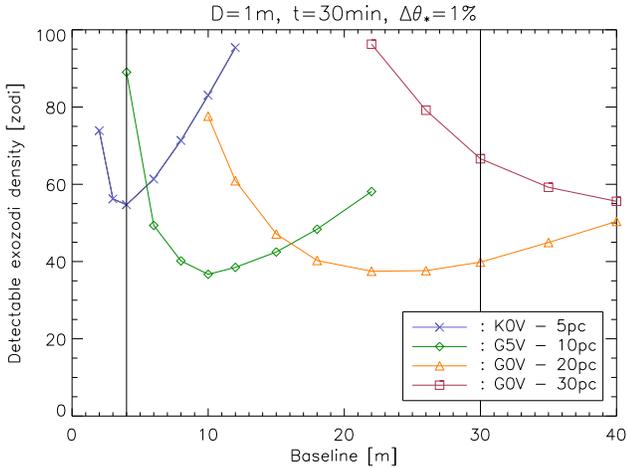}} \caption{Simulated performance of
ALADDIN in terms of exozodiacal disc detection for four typical Darwin targets. The baseline design
for ALADDIN has been used (see Table~\ref{tab:instruparam}), assuming an integration time of
30~min, an uncertainty of 1\% on the knowledge of the stellar angular diameter, and using the
worst-case scenario for atmospheric turbulence (see text). The two vertical lines indicate the
baseline range proposed for the ALADDIN conceptual design.} \label{fig:perfbase}
\end{figure}

In Fig.~\ref{fig:perfbase}, we present the results of the simulations in terms of detectable
exozodiacal density level as a function of baseline length. As in \citetalias{Absil06a}, the
threshold for detection is set at a global signal-to-noise of 5, including the residuals from
background subtraction and from geometric and instrumental stellar leakage calibration.
Fig.~\ref{fig:perfbase} shows that the optimum baseline for studying typical Darwin target stars is
comprised between about 4 and 40~m, which closely matches the baseline range offered by ALADDIN.
With its 1~m class telescopes, ALADDIN significantly outperforms GENIE for the same integration
time in the case of nearby targets (see Table~\ref{tab:aladdingenie} for a thorough comparison).
This fact is not only due to the exceptional atmospheric conditions, but also to the optimisation
of ALADDIN both regarding the available baselines and the instrumental design.

To check the relevance of ALADDIN in the context of the Darwin preparatory science, it is useful to
compare the angular resolution provided by the optimum baseline length with respect to the position
of the habitable zone for the various targets, because Darwin will be most sensitive to dust
located in that particular zone where it will search for Earth-like planets. According to
\citet{Kasting93}, the position of the habitable zone expressed in AU is given in good
approximation by the following equation:
\begin{equation}
r_{\rm HZ} = \left( \frac{T_{\star}}{T_{\odot}} \right)^2 \frac{R_{\star}}{R_{\odot}} \; ,
\end{equation}
which yields 0.68, 0.85 and 1.16~AU for a K0V, a G5V and a G0V star respectively. The angular
distance of the habitable zone to its parent star is compared to the angular resolution of ALADDIN
in Table~\ref{tab:optbase}. The first bright fringe of the optimised nulling interferometer always
falls between the star and the habitable zone, and the associated angular resolution is compatible
with the study of this most important region of the exozodiacal disc. This also validates {\em a
posteriori} the choice of the $L$ band for the study of exozodiacal dust around the Darwin target
stars.

\begin{table}[t]
\begin{center}
\caption{Comparison of the angular resolution provided by the optimum ALADDIN baseline length ($4
\le b \le 30$\,m) with the characteristic position of the habitable zone of the target systems.}
\label{tab:optbase}
\begin{tabular}[h]{cccc}
\hline      & Optimum  &  Ang. resol.   & Position HZ
\\          & baseline & $(\lambda/2b)$ & $(r_{\rm HZ}/d)$
\\ \hline K0V 5~pc  & 4 m  & 93 mas  & 135 mas
\\        G5V 10~pc & 10 m & 37 mas  & 85 mas
\\        G0V 20~pc & 24 m & 15 mas  & 58 mas
\\        G0V 30~pc & 30 m & 12 mas  & 39 mas
\\ \hline
\end{tabular}
\end{center}
\end{table}

    \subsection{Calibration of stellar angular diameters}

\begin{figure}[t]
\centering \resizebox{\hsize}{!}{\includegraphics{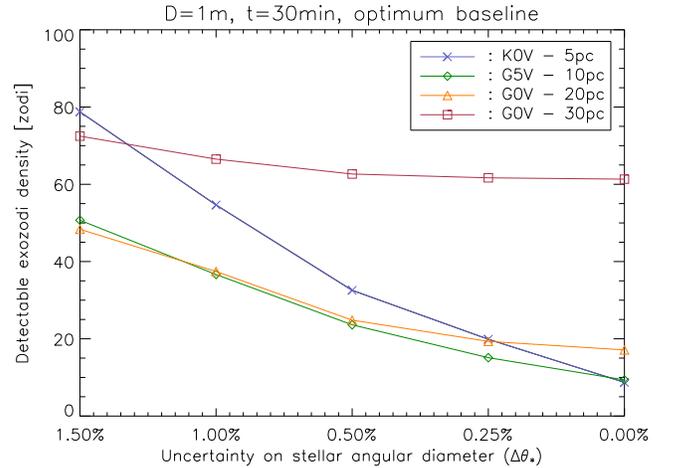}} \caption{Simulated performance of
ALADDIN in terms of exozodiacal disc detection for various assumptions on the uncertainty on the
stellar angular diameters.} \label{fig:perfang}
\end{figure}

An important parameter influencing the performance of a nulling interferometer is the uncertainty
on the angular diameter of the target star ($\Delta \theta_{\star}$). It is the main contributor to
the quality of calibration not only for geometric stellar leakage but also for instrumental stellar
leakage, which relies on the estimation of the instrumental nulling ratio on a well-known
calibration star. In Fig.~\ref{fig:perfang}, we investigate the influence of this $\Delta
\theta_{\star}$ parameter on the ALADDIN sensitivity. The baseline length is optimised in each case
within the specified range ($4-30$~m). This simulation shows that, similarly to the GENIE case, an
improved accuracy on stellar diameters would largely improve the detection capabilities of ALADDIN.

The very good sensitivity obtained for a perfect knowledge of the stellar diameter gives an idea of
the gain that could be achieved by using more elaborate nulling configurations that are almost
insensitive to stellar leakage. An example of such a configuration is the Degenerate Angel Cross
\citep{Mennesson05}, which uses three aligned telescopes to provide a central transmission
proportional to the fourth power of the angular distance to the optical axis ($\theta^4$) instead
of the second power ($\theta^2$) for a two-telescope Bracewell interferometer. The use of phase
chopping with multi-telescope configurations would have almost the same effect, as geometric
stellar leakage would then be removed by the chopping process. Fig.~\ref{fig:perfang} shows that an
advanced nulling interferometer at Dome~C should be capable of reaching a sensitivity ranging
between 10 and 20~zodi around most of the Darwin targets. Multi-telescope configurations are
however not contemplated in the context of ALADDIN, for which simplicity is strongly advocated.

\begin{table}[t]
\begin{center}
\caption{Comparison of the GENIE and ALADDIN performance expressed in detectable exozodiacal disc
densities as compared to the solar zodiacal disc. Four different levels of uncertainty have been
assumed on the angular diameter of the target stars. The simulations are performed in the $L$ band,
which extends from 3.5 to 4.1~$\mu$m in the case of GENIE and from 3.1 to 4.1~$\mu$m in the case of
ALADDIN. An integration time of 30~min is assumed in all cases.} \label{tab:aladdingenie}
\begin{tabular}[h]{c|cccc|c}
\hline       Star     &0.25\%&0.5\%&1\% &1.5\%& Instrument
\\ \hline             &  72 &  90 & 125 & 154 & GENIE -- AT
\\        K0V -- 5pc  & 114 & 227 & 455 & 682 & GENIE -- UT
\\                    &  20 &  33 &  55 &  79 & ALADDIN
\\ \hline             & 111 & 130 & 154 & 176 & GENIE -- AT
\\        G5V -- 10pc &  30 &  59 & 117 & 176 & GENIE -- UT
\\                    &  15 &  24 &  37 &  51 & ALADDIN
\\ \hline             & 255 & 261 & 278 & 297 & GENIE -- AT
\\        G0V -- 20pc &  21 &  29 &  50 &  73 & GENIE -- UT
\\                    &  19 &  25 &  37 &  48 & ALADDIN
\\ \hline             & 575 & 585 & 604 & 615 & GENIE -- AT
\\        G0V -- 30pc &  36 &  46 &  59 &  71 & GENIE -- UT
\\                    &  62 &  63 &  67 &  72 & ALADDIN
\\ \hline
\end{tabular}
\end{center}
\end{table}

Table~\ref{tab:aladdingenie} compares the expected sensitivity of ALADDIN, operated on 1-m
telescopes, with that of GENIE on either 8-m Unit Telescopes or 1.8-m Auxiliary Telescopes at the
VLTI, using various assumptions on the stellar diameter knowledge. A significant gain (up to a
factor~4) is obtained with ALADDIN, except in the case of the G0V at 30~pc where the 8-m telescopes
are providing a more suited collecting area. The gain with ALADDIN is all the larger when the
target star is closer, because the use of short baselines is crucial for stars with relatively
large angular diameters ($\gtrsim$1~mas). As obvious from Table~\ref{tab:aladdingenie}, an accurate
knowledge of the stellar angular diameter ($<0.5$\%) at the observing wavelength is mandatory to
reach our goal sensitivity of 20~zodi.

\begin{figure}[t]
\centering \resizebox{\hsize}{!}{\includegraphics{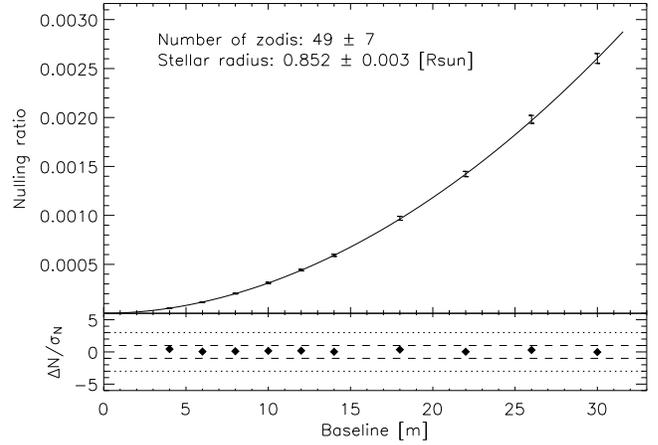}} \caption{Simultaneous fit of the
stellar radius and the exozodiacal density level using ten 30-min observations of a K0V star at
5~pc (actual radius of $0.85 R_{\odot}$), surrounded by a 50-zodi disc. The nulling ratio is
computed as the ratio between the destructive and constructive outputs of the beam combiner. The
contribution of instrumental stellar leakage to the nulling ratio has been subtracted from this
data using the same procedure than elaborated for GENIE: the observation of a calibration target of
similar type and distance, but without circumstellar material.} \label{fig:fit}
\end{figure}

Angular diameters in the $L$ band are however currently not well constrained, due to the lack of
actual measurements. Furthermore, it is not guaranteed that an interferometer will operate in this
band in a near future to provide angular diameter measurements with the required accuracy, while
extrapolating stellar models from the visible or near-infrared ($H$, $K$ bands) towards the $L$
band is not straightforward \citepalias[see][]{Absil06a}. An integrated concept as ALADDIN presents
a significant advantage in this respect, as the continuous range of available baselines can be used
to fit the stellar angular diameter simultaneously with the exozodiacal disc parameters. This
procedure is illustrated in Fig.~\ref{fig:fit}, where we have simulated ten 30-min observations of
a K0V star at 5~pc surrounded by a 50-zodi disc, using ten baseline lengths ranging between 4 and
30~m. All standard calibrations have been applied in these observations, except for the calibration
of the geometric stellar leakage (which is actually the main contributor to the observed nulling
ratio). The simultaneous fit of the stellar radius and the exozodiacal dust density level provides
encouraging results, and confirms that the {\em a priori} knowledge of the stellar radius is not
required if a sufficient baseline coverage is used.

    \subsection{Influence of integration time}

Another advantage of the ALADDIN concept is its ability to perform very long on-source
integrations. The interferometer is assumed to be continuously operated during the long winter
night, but also during the equinox twilight and the summer day thanks to the low sky temperature in
all seasons and to the very low aerosol and dust content in the atmosphere (coronal sky). The
summertime performance will of course be somewhat degraded due to the unavoidable stray light in
the instrument and to the higher temperature of the sky and optical train, which produces a larger
background emission. Long integrations are also enabled by the fact that ALADDIN would be dedicated
to the survey of exozodiacal discs, while an instrument like GENIE would have to compete with other
instruments at the VLT (especially when using the 8-m Unit Telescopes). Therefore, it makes sense
to investigate the gain in sensitivity that can be achieved by longer integrations. The computation
of this gain is not trivial, as all the noise sources do not have the same temporal behaviour. For
instance, shot noise, detector noise and instability noise (to the first order) have the classical
$t^{1/2}$ dependence, while the imperfect calibration of geometric and instrumental stellar leakage
is proportional to time (it actually acts as a bias).

\begin{figure}[t]
\centering \resizebox{\hsize}{!}{\includegraphics{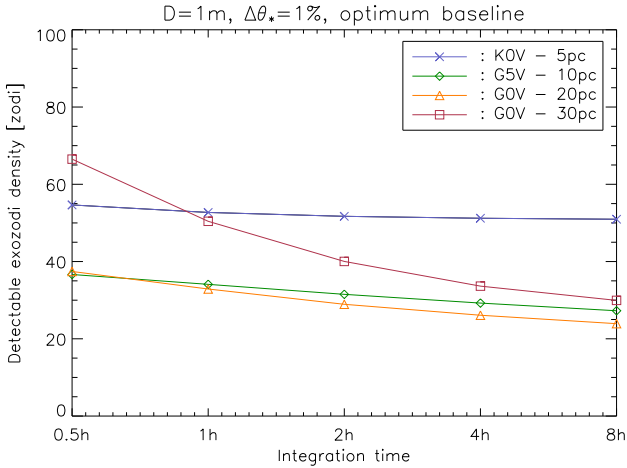}} \caption{Simulated performance of
ALADDIN in terms of exozodiacal disc detection for increasing integration times.}
\label{fig:perftime}
\end{figure}

In Fig.~\ref{fig:perftime}, we simulate the sensitivity of ALADDIN as a function of integration
time. Because increasing the integration time does not improve the accuracy of both geometric and
instrumental stellar leakage calibration, which are among of the main contributors to the noise
budget (especially for very nearby stars), the overall performance does not largely improve for
long exposures (except for the fainter targets, for which sensitivity is background-limited). It
must still be noted that the goal sensitivity of 20~zodi is within reach after 8~hours of
integration for G0V stars located closer than 20~pc.

A side effect of increasing the integration time is that the optimum baseline is decreased. Indeed,
shorter baselines allow for less exozodiacal light to make it through the transmission pattern, but
also for a better cancellation of the stellar light. The result is an improved signal-to-noise
ratio regarding stellar leakage calibration, while the relative increase of the shot noise
contribution with respect to the transmitted exozodiacal signal is compensated by the longer
integration time. For instance, in the case of a G0V star at 20~pc, the optimum baseline decreases
from 24~m for a 30~min integration to 12~m for an 8~h integration. Reducing the optimum baseline is
favourable to the global feasibility of the concept, as it reduces the required size of the truss
supporting the siderostats.

    \subsection{Influence of pupil diameter}

Finally, in order to choose the most appropriate diameter for the ALADDIN siderostats, we study the
influence of the collecting area on the sensitivity of the instrument. To keep the system
architecture unchanged, we restrict the pupil diameter to 2~m at most, since larger pupils would
become comparable to the size of turbulent cells above the boundary layer (about 4~m in the $L$
band) and would therefore require either adaptive optics or additional intensity control to be
implemented.

\begin{figure}[t]
\centering \resizebox{\hsize}{!}{\includegraphics{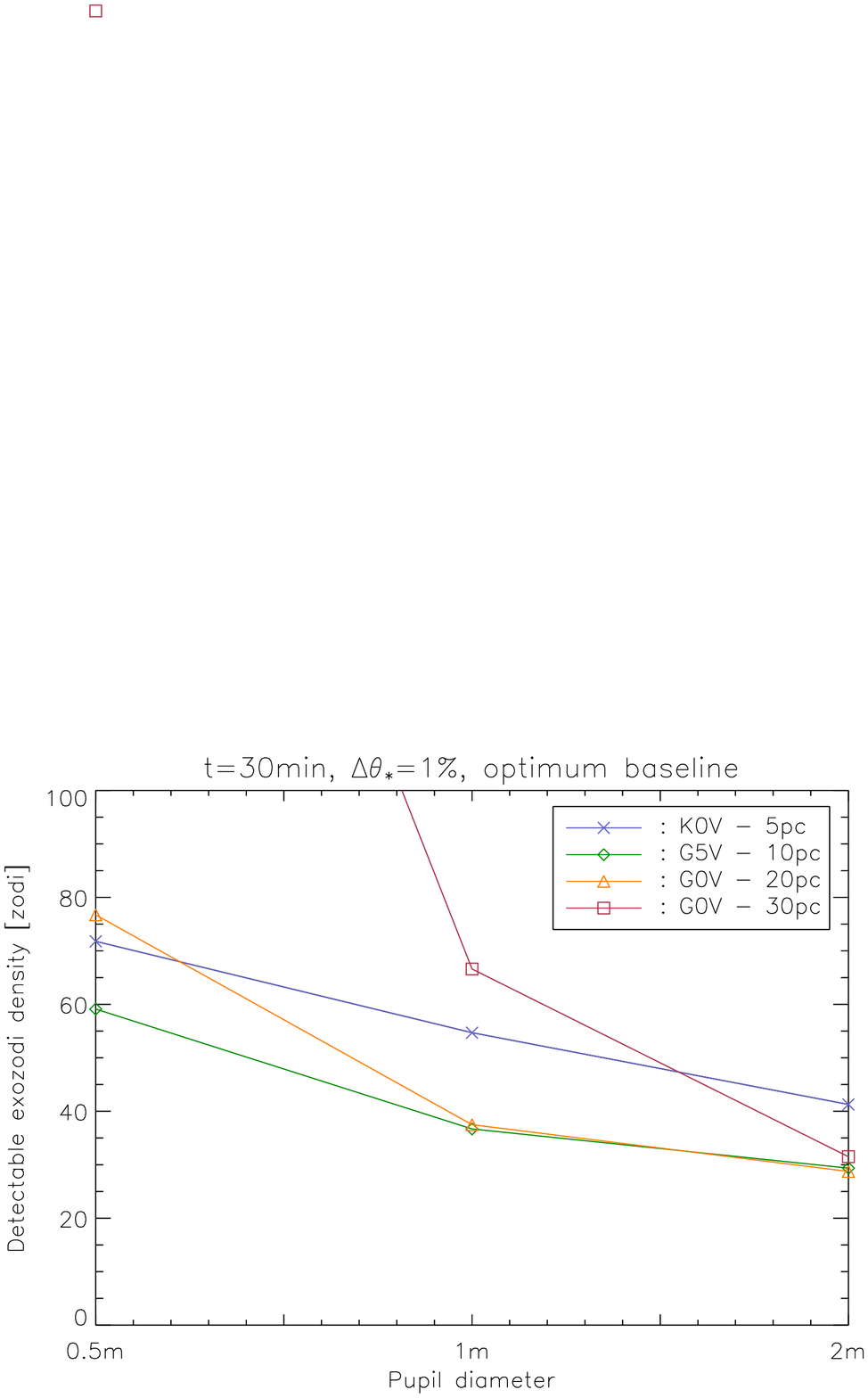}} \caption{Simulated performance of
ALADDIN in terms of exozodiacal disc detection for three different pupil diameters (50~cm, 1~m and
2~m).} \label{fig:perfdiam}
\end{figure}

Fig.~\ref{fig:perfdiam} shows the simulated performance of ALADDIN for three different sizes of the
siderostats. By increasing the diameter from 0.5~m to 1~m, the performance improves by a factor
ranging from 25\% to 75\%, while a typical gain between 25\% and 50\% is observed when increasing
the telescope size from 1~m to 2~m. The G0V star at 30~pc shows the most significant improvement as
a function of pupil size, due to its faintness (shot noise from the background emission is dominant
for such a faint star). It must be noted that the performance of ALADDIN with 50-cm collectors is
still better than that of GENIE at the VLTI for the two closest targets. Reducing the size of the
siderostats could thus make sense if the feasibility of the project was found to be jeopardised by
the requirement to put 1-m siderostats on a 40-m truss located 30~m above the ground. Increasing
the integration times by a factor about~4 would then be required to achieve similar performance as
with 1-m collectors. A beneficial side-effect of increasing the integration time will be to reduce
the optimum baselines down to an acceptable length, because 50-cm siderostats are associated with
optimum baselines typically twice as large as for the original 1-m collectors. In practice, the
final choice of the pupil diameter will result from a trade-off between feasibility, performance,
integration time and available baselines.

\section{Site impact on performance}

In this section, we estimate the gain in performance that is actually related to the outstanding
observing conditions above the boundary layer on the Antarctic plateau (and not to the optimised
instrumental design). For that purpose, we simulate the performance of ALADDIN at two other
locations: first on the ground at Dome~C (below the boundary layer) and then at Cerro Paranal. In
the first case, we use the ground-level wintertime seeing conditions recently reported at Dome~C by
\citet{Agabi06}: a median seeing of $1\farcs9$ (i.e., a Fried parameter of 5.4~cm at 500~nm) and a
coherence time of about 2.9~msec (i.e., equivalent wind speed of 5.8~m/s). In the second case, we
use the standard atmospheric conditions of Cerro Paranal already presented in
\citetalias{Absil06a}.

    \subsection{Ground-level performance at Dome C}

One of the main limitations of ground-level observations comes from the fact that the Fried
parameter in the $L$ band ($\sim 57$~cm) becomes smaller than the size of the apertures, so that
multiple speckles are formed in the image plane. Assuming only tip-tilt control at 1~kHz, which
provides a residual tip-tilt of about 15~mas, the typical fluctuations of the relative intensity
mismatch between the two beams after modal filtering would be about 18\%. This is much too large to
ensure a high and stable instrumental nulling ratio, and the use of adaptive optics (or of an
intensity matching device) is therefore required to stabilise the injection into the single mode
waveguides. Another limitation comes from the increased strength of the piston effect. Assuming
fringe tracking to be performed at a maximum frequency of 10~kHz, the residual OPD would range
between 15 and 35~nm rms depending on the target star. Here again, the stability of the nulling
ratio would be significantly degraded with respect to the baseline ALADDIN concept. On the
contrary, longitudinal dispersion is not expected to increase very significantly since the
precipitable water vapour content of the first 30~m of the atmosphere is relatively small due to
the very low temperature right above the ice.

Taking all these effects into account, the instability of the nulling ratio ({\em instability
noise}) would become the main source of noise in the budget of a ground-level ALADDIN. The
simulations performed with GENIEsim show that the sensitivity in the case of a G0V star located at
20~pc would only be about 200~zodi instead of 37~zodi for the original ALADDIN concept on top of a
30-m tower. In order to match the baseline ALADDIN performance with a ground-level instrument,
higher repetition frequencies would be required for piston and tip-tilt control (both about 6~kHz),
while adaptive optics (or intensity control) should be used to stabilise the injection efficiency
into the waveguides. Dispersion control might also be required. Preliminary estimations show that
deformable mirrors using $20 \times 20$ actuators at a repetition frequency around 1~kHz would be
required to reduce the intensity fluctuations down to 1\%. In that case, a sensitivity around
50~zodi would be reachable for a G0V star at 20~pc.

Obviously, placing the instrument above the ground layer is recommended to obtain a significant
gain on both the performance and feasibility aspects with respect to an instrument installed at a
temperate site such as Cerro Paranal.

    \subsection{Ground-level performance at Cerro Paranal}

To better emphasise the attractiveness of Antarctic sites in the context of high dynamic range
interferometry, let us now virtually move the ALADDIN experiment to Cerro Paranal while keeping the
design unchanged. Because the Fried parameter is larger at Paranal ($r_0 \sim 1.2$~m in the $L$
band) than at the ground level at Dome~C, while the coherence time is of the same order of
magnitude ($\tau_0 \sim 3$~msec), the performance should be somewhat better than on the ground at
Dome~C. Simulations indeed show that the residual OPD is slightly improved (now between 10 and
30~nm), while the residual intensity fluctuation is significantly reduced (now about 7\%, but still
well above the goal of 1\%).

However, two other parameters significantly degrade the situation: the large background emission
and the increased PWV content in the atmosphere. The main effect of the former is to increase the
integration time to reach a given sensitivity limit, while the fluctuations of the latter produce
large variations of longitudinal dispersion, which can reach about 0.7~radian if they are not
reduced by a real-time control loop as in the case of GENIE \citepalias[see][]{Absil06a}. This
corresponds to an additional OPD error of about 400~nm at the edges of the observing waveband
(ranging from 3.5 to 4.1~$\mu$m in the case of Cerro Paranal). All in all, a sensitivity of about
3000~zodi is expected for a replica of ALADDIN installed at Cerro Paranal. By introducing a
dispersion control loop similar to that described in \citetalias{Absil06a} and operating it at a
typical frequency of 50~Hz, longitudinal dispersion could be reduced down to about 0.05~radian
(30~nm), in which case the sensitivity would reach about 250~zodi. This would however significantly
increase the technical complexity of the instrument, which is not desired.


\section{Conclusion} \label{sec:conclusion}

In this paper, we have investigated a potential solution to a well-defined scientific need, viz.\
characterising the dusty environment of candidate target stars for future life-finding missions
such as Darwin or TPF. In a previous study \citepalias{Absil06a}, we have shown that an infrared
nulling interferometer installed on a temperate site, such as the GENIE project at Cerro Paranal,
would provide useful information on candidate targets, but (1) that its technical feasibility could
be jeopardised by the requirement to design complicated control loops for mitigating the effects of
atmospheric turbulence, and (2) that its sensitivity would not reach the desired level of 20~times
the density of our local zodiacal cloud.

To overcome these two limitations, we propose in this paper a conceptual design for a nulling
interferometer (ALADDIN) to be installed at Dome~C, on the high Antarctic plateau. Based on the
atmospheric turbulence measurements obtained so far at Dome~C, we have updated the GENIEsim
software \citepalias{Absil06a} to simulate the performance of such an instrument. These simulations
show that, using 1-m collectors, this instrument would have an improved sensitivity with respect to
GENIE working on 8-m telescopes, provided that it is placed above the turbulence boundary layer,
which is about 30~m thick at Dome~C. In particular, the 20-zodi sensitivity goal seems within reach
for typical Darwin/TPF target stars. Moreover, the exceptional turbulence conditions above the
boundary layer significantly relax the requirements on the real-time compensation of atmospheric
effects, improving the feasibility of the instrument. It must also be noted that, thanks to the
optimised range of adjustable baselines, the harmful influence of our imperfect knowledge of
stellar angular diameters can be largely mitigated by simultaneously fitting a photospheric model
and an exozodiacal disc model to the collected data, yet at the price of an increased observing
time.

While we assumed the instrument would be deployed above the boundary layer at Dome~C, site of the
Concordia station, it might turn out that other sites on the Antarctic plateau provide
simultaneously a thinner boundary layer and an improved free air seeing, and hence better
performance. The final choice for the site will have to trade off the practical advantages of
feasibility and performance vs.\ logistical support.

This paper illustrates the potential of Antarctic sites for high-angular, high-dynamic range
astrophysics in the infrared domain. In the particular, well specified case of a nulling
interferometer, we were able to realistically quantify the relative gain with respect to a
temperate site, showing that a pair of 1\,m telescopes on the plateau will perform better than a
pair of 8\,m telescopes at Cerro Paranal. Other applications would result in different gains, but
it is clear that there are niches where the Antarctic plateau enables observations that would
otherwise require access to space.


\begin{acknowledgements}
The authors are indebted to R.~den Hartog and D.~Defr\`ere for their major contributions to the
development of the GENIEsim software, which has been used throughout this paper. The authors also
wish to thank the engineers at Thales Alenia Space that have contributed to the preliminary design
of the ALADDIN instrument, as well as T.~Fusco for the simulation of adaptive optics performance at
Dome~C. O.A.~acknowledges the financial support of the Belgian National Fund for Scientific
Research (FNRS) while at IAGL and of a Marie Curie Intra-European Fellowship (EIF) while at LAOG.
\end{acknowledgements}


\bibliographystyle{aa} 
\bibliography{7582} 

\end{document}